\documentclass[doublecol]{epl2}
\usepackage{amsmath}
\usepackage{amssymb}
\usepackage{amsfonts}
\usepackage{graphicx}
\usepackage{dcolumn}
\usepackage{bm}
\usepackage{sidecap}
\usepackage {caption}
\usepackage{float}
\usepackage{parskip}

\title{Optimization of synchronizability in multiplex networks}

\author{Sanjiv K. Dwivedi$^{1}$, Camellia Sarkar$^{2}$ and Sarika Jalan$^{1,2,}$\footnote{Corresponding Author: sarikajalan9@gmail.com}}
\institute{$^1$Complex Systems Lab, Physics Discipline, Indian Institute of Technology Indore,
Khandwa Road, Indore-452017\\ $^2$Centre for Biosciences and Biomedical Engineering, Indian Institute of Technology Indore,
Khandwa Road, Indore-452017 }

\abstract
{We investigate the optimization of synchronizability in multiplex networks and demonstrate that the interlayer coupling strength is the deciding factor for the efficiency of optimization. 
The optimized networks have homogeneity
in the degree as well as in the betweenness centrality.
Additionally, the interlayer coupling strength crucially affects various properties of 
individual layers in the optimized multiplex networks. We provide an understanding to 
how the emerged network properties are shaped or affected when the evolution 
renders them better synchronizable.
}

\begin{document}
\maketitle
{\bf Introduction:} Synchronization is one of the most fascinating phenomena witnessed in interacting dynamical units \cite{book_kurths}. 
Given a wide applicability of this phenomenon, in the last two decades, a large number of studies have been devoted to understand the interplay of the individual dynamical units and the underlying network structure
\cite{Boccaletti}, most importantly how structural properties affect the
synchronizability \cite{Arenas,Igor_PRL_2005}. 
Furthermore, various spectral properties of the underlying networks
impart an understanding to the collective dynamical behavior \cite{Motter,Bogu,Restrepo1,Nishikawa},
with the extremum eigenvalues of the corresponding
Laplacian matrices providing insight in to the synchronizability of a network \cite{Barahona,Restrepo,Pecora}.

So far most of the works pertaining to the synchronization as well as optimally synchronized networks are restricted to the single layer networks, where attempts have been made to understand the evolutionary origin of the most synchronizable networks based on the minimization of the ratio (R) of the largest and the first nonzero eigenvalues of the Laplacian matrices using simulated annealing \cite{Luca}. However, multilayer or multiplex networks have been increasingly realized to present a more realistic framework to model the interactions in real
world systems \cite{BoccalettiRep}.
Given the importance of the multilayer framework, there is a spurt in the activities of understanding and characterizing their structural \cite{Battiston} and spectral properties \cite{Gomez,Ribalta}.
These works have shown that the synchronizability
of multilayer networks is dependent on the interlayer coupling strength
and attains a maximum value at some critical value of the strength, 
however, the degree of efficiency of optimization is not understood with respect to the interlayer coupling strength \cite{Ribalta}.
Furthermore, a weighted nature
of the coupling strength has been realized in many real world networks 
\cite{weighted1,weighted2}. The optimization of synchronizability in
weighted coupling is a complex problem due to the contribution arising from the degree of freedom in the coupling strength as well as the connection pattern of the networks.
Few works pertaining to understand the impact of weighted nature of coupling strength on 
the synchronizability of a network conclude
that the scale-free networks are more sensitive to various optimization methods \cite{Chung_book,Atay,Zhou,Chavez}. 

In this Letter, we investigate the impact of structural properties of a network on its synchronizability under the multiplex network framework. The multiplex networks are 
evolved using the Genetic algorithm (GA) considering $R$ as the fitness function. We find
that the best synchronizable multilayer network has a stronger interlayer connectivity as
compared to the connections within each layer. Further, the optimal network supports homogeneity 
in the degree, in addition to an emergence of the degree-degree correlations for the mirror nodes 
in different layers, reflecting several distinguished features incorporating multiplexity.
We provide an understanding of various emerging structural features of the final evolved
networks by relating them 
with the load distribution on the interlayer links
\cite{Nishikawa} as well as by 
the spectral properties of the underlying Laplacian matrices \cite{Chung_book,Ribalta}. 
\begin{figure}[t]
\centerline{\includegraphics[height= 3cm,width = 8cm]{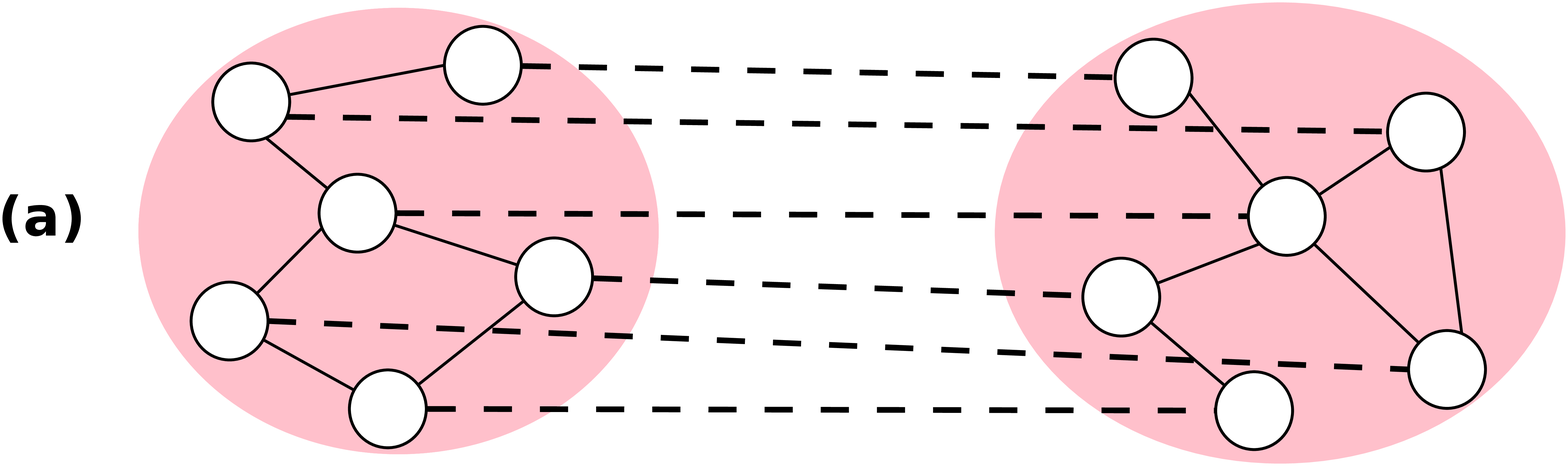}}
\centerline{\includegraphics[height= 1.7cm,width = 8cm]{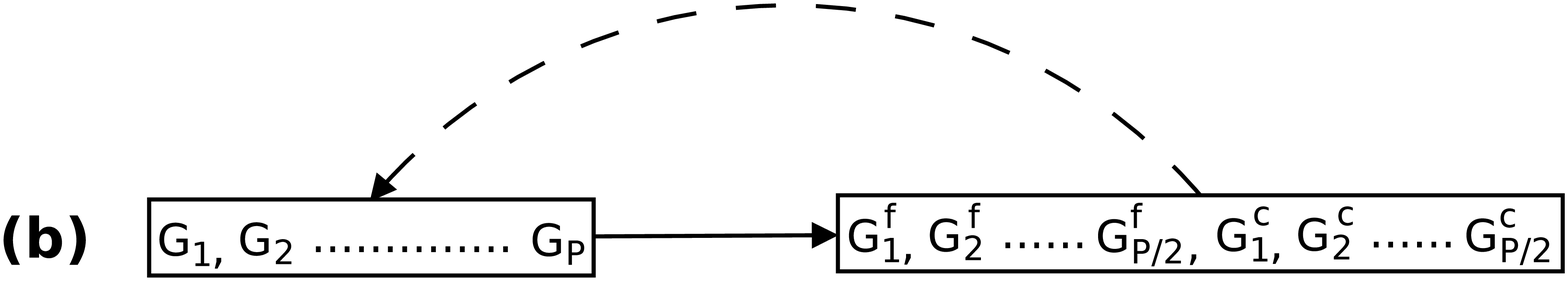}}
\caption{(a) Schematic diagram of a multiplex network consisting of two layers
in which the solid lines represent intralayer couplings, whereas the
dashed lines depict interlayer couplings.
(b) Schematic diagram denoting evolution of networks using the genetic algorithm,
where $G_{i}$ represents the $i^{th}$ network in a population.
$G^f_{i}$ and $G^c_{i}$ represent $i^{th}$ fitter parent and child networks, respectively.
The solid arrow represents progression of generation from the initial population and the dashed arrow indicates that the fitter and child graphs form the initial population for the next generation.}
\label{Multiplex}
\end{figure}
\begin{figure}[h]
\centerline{\includegraphics[width=1\columnwidth]{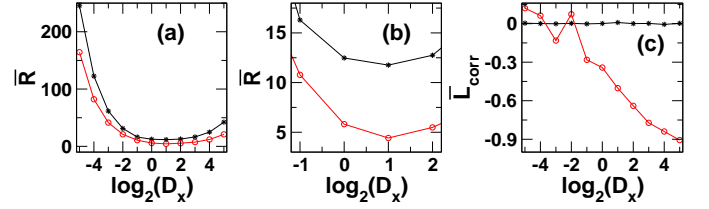}}
\caption{Minimized (circles) and initial (stars) average value of $R$ for various interlayer 
coupling strength in panel (a) and its zoomed version in panel (b).
Panel (c) shows the values of the correlation $\overline{L}_{\mathrm{corr}}$ for initial ER random networks (stars) and optimized networks (circles).
For each case, GA minimize the $R$ till 6000 iterations and the average is taken over population of 1000 networks and for the last 1000 iterations. Each initial layer of ER networks have $\langle k\rangle$=6 and $N=100$ \cite{size_impact}.}
\label{FigERAv6}
\end{figure}

{\bf Theoretical framework:} We consider a bi-layer multiplex network, each layer having $P$ 
network population with a fixed size $N$
and a fixed average degree $\langle k \rangle$. The networks representing different layers may have
different architecture. Let the adjacency matrix of the first and second layers be denoted by 
{$A^{i}$}
and {$B^{i}$}, respectively, where $1\leqslant i \leqslant P$.
The weighted adjacency matrix $M^{i}$ of size 2N$\times$2N for the multiplex network
(Fig.~\ref{Multiplex}(a))
is defined by
\begin{equation}
 M^i = \left[
    \begin{array}{cc}
      A^i~~D_{x}I\\
      D_{x}I~~B^i
    \end{array}
\right] 
\label{rec_mat}
\end{equation}
where I is the identity matrix of size N$\times$N and $D_{x}$ is the interlayer coupling strength.
\begin{figure}[t]
\centerline{\includegraphics[width=\columnwidth]{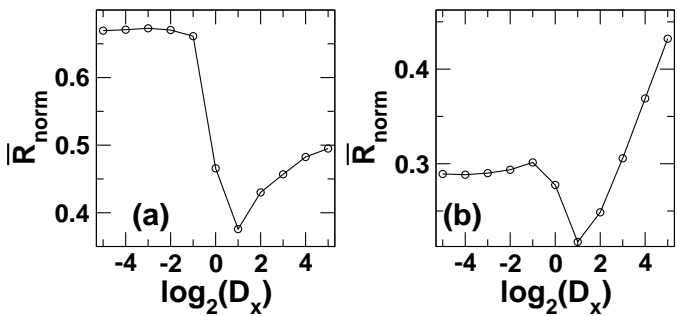}}
\caption{Average value of $R_{norm}$ when individual network in a population is represented as (a) ER and
(b) scale-free networks, respectively. For each case, GA minimizes the value of $R$
till 6000 iterations and the average is taken over the population of 1000 networks
for last 1000 iterations. Each layer of the networks have $\langle k\rangle$=6 and $N=100$,
which are preserved throughout the evolution.}
\label{FigERAv6Norm}
\end{figure}

The Laplacian matrix $L^{i}$ obtained from $M^{i}$ is defined by
\begin{equation}
L^{i}_{jk} = \begin{cases} ~~~d^i_{j}~~~~\mbox{if } {~~~ j\not=k }\\
- M^{i}_{jk} ~~ \mbox{otherwise}, \end{cases}
\label{rec_mat1}
\end{equation}
where $d^i_j = \sum_k M^i_{jk}$.
Let $\lambda^{i}_{1}$ $<$ $\lambda^{i}_{2}$ $\leqslant$ .... $\leqslant$ $\lambda^{i}_{2N}$
be the eigenvalues of $L^i$.
The eigenvalue ratio $R^i$ = $\lambda^{i}_{2N}$/$\lambda^{i}_{2}$ of the $L^i$ matrix
is used as a fitness function of $i^{th}$ multiplex network in the genetic algorithm. 
A network having the lesser $R$ value is  more fit. Out of $P$ population of the
initial networks, we select $P/2$ fitter networks for the next generation, which we term as the parent networks (Fig.~\ref{Multiplex}(b)).  
The remaining $P/2$ networks of the next generation, known as the child networks, are 
constructed from the parent networks as follows. First a pair of the matrices is
selected randomly from the $P/2$ population and then the adjacency matrices of these
selected parent networks, termed as the parent matrices,
are used to construct one crossed child matrix. Each block of this child matrix 
is constructed from the block having the same position in one of the parent matrices which
are selected with 
the equal probability.
Then, by taking the upper triangular part of the crossed matrix, a symmetric child matrix is
built leading to the undirected child network.
Further, the average degree of the crossed matrix is preserved by randomly
inserting or deleting 1 and 0 entries decided by a probability. This probability is dependent on 
the fluctuations of $1$ entries from the adjacency matrices of the parent networks. The details of these steps are given
in \cite{GAclstering}.
The process is repeated until we generate $P/2$ child networks which along with the $P/2$
parent networks form the 
 same size of the population as of the initial networks.
The networks are evolved until we attain the steady state.
We measure various structural properties of networks as they evolve
and analyze the interplay of these properties with the
optimization of synchronizability.
\begin{figure}[h]
\centerline{\includegraphics[width=1\columnwidth]{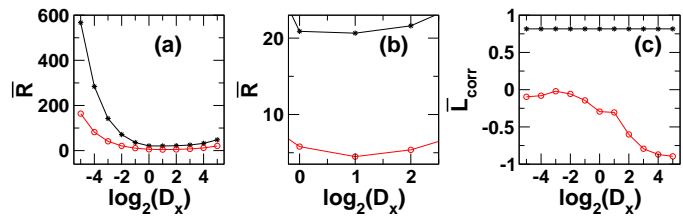}}
\caption{Minimized (circles) and initial (stars) average value of $R$ for various interlayer coupling strength in 
panel (a)
and its zoomed version in panel (b).
Panel (c) shows the values of the correlation $\overline{L}_{\mathrm{corr}}$ for initial SF networks (stars) and optimized networks (circles).
For each case GA minimize the $R$ till 6000 iterations and the average of is taken over population of 1000 networks
and last 1000 iterations. Each initial layer of scale-free networks have $\langle k\rangle$=6 and $N=100$.}
\label{FigSFAv6}
\end{figure}

{\bf Results and Discussion:} In order to draw a close coherence with the real world networks, we consider the sparse networks as the initial population in GA.
Starting with networks having random architecture, we evolve them using GA taking $R$ as the fitness function.
We find that at the lower interlayer couplings ($D_{x} << 1$), the $\overline{R}$ is  
high which 
decreases with an increase in the $D_{x}$ values.
At a certain $D_{x}$ value, the $\overline{R}$ attains its minima, then saturates followed by an increase in its value for $D_{x} >> 1$ (Fig.~\ref{FigERAv6}(a),(b)). 
Note that for all the simulations, the size and the average degree of individual layer 
remains same as for the initial network, hence, any change in the value of
$\overline{R}$ for initial network population  
is arising due to changes in the $D_{x}$ values.
The lower $D_{x}$ values indicate the higher values of $R$ as at 
lower interlayer coupling strength the $\lambda_{2}$ values are low and at higher $D_{x}$ values 
the $\lambda_{N}$ values are high. Both the factors contribute to an increase in the
$R$ values \cite{Ribalta} and for the model considered in the Letter,
$R$ can be determined as following: 
Case(1), when $D_{x} << 1$
\begin{equation}
 R \approx \frac{\max\limits_{\alpha}\big[\lambda_{max}(L^{\alpha}) ~+~D_{x}\big]}{2 D_{x}}
\label{EQ1}
\end{equation}
 where $L^\alpha$ is the Laplacian of $\alpha^{th}$ layer, $\lambda_{max} (L^\alpha)$ is the maximum eigenvalue of the Laplacian of $\alpha$th layer.
Case(2), when $D_{x} >>$ 1
\begin{equation}
 R \approx \frac{2 D_x + \sqrt{2} \lambda_{max}(L^{AV})}{\lambda_2(L^{AV})}
\label{EQ2}
\end{equation}
where $L^{AV}$ is average Laplacian of two layers \cite{Ribalta}.
Eq.~\ref{EQ1} indicates that $R$ is a decreasing function of $D_{x}$ for
weaker interlayer coupling strength
whereas Eq.~\ref{EQ2} indicates that $R$ is an
increasing function of $D_{x}$ for the stronger interlayer coupling strength. Consequently there is a trade off between both 
the behaviors leading to a value of $D_{x}$ for which $R$ achieves its minimum value. It turns out that this minimum
value occurs at $D_{x}$ = 2 and hence Eq.~\ref{EQ2} stands as a better approximation for $R$. 
What follows that the structural changes in
the architecture of the optimized networks affect the second term in the numerator and the denominator which collectively
contribute to minimization of the $R$ values.

Further, we define the efficiency of optimization as $R_{norm} = R_{opt}/R_{ini}$, where
$R_{ini}$ and $R_{opt}$ are the eigenvalue ratio of the initial and the optimized networks, 
respectively.  A lower value of $R_{norm}$ reflects a high efficiency of the synchronizability.
 For the ER random networks, initially with an increase in the $D_{x}$ values, 
$R_{norm}$ remains almost constant until the interlayer coupling strength reaches to 
0.50 (Fig.~\ref{FigERAv6Norm}(a)). However, the value of $R_{norm}$ being less than one implies 
that there is an optimization efficiency with respect to the
initial population. Thereafter, it exhibits an abrupt decrease and attains a minimum value for 
$D_{x}$ = 2. Therefore, GA finds a variability in the efficiency of the 
optimization for various values of $D_{x}$.  
The value of $R_{norm}$ starts increasing after attaining the minimum value.
The above results indicate that the enhancement in the first term of the numerator of
Eq.~\ref{EQ2}, which is proportional to the interlayer coupling strength, might lead to a decrease in the efficiency of the optimization ($R_{norm})$. 
The GA, thus maximizes the synchronizability for various values of the interlayer
coupling strength (Fig.~\ref{FigERAv6}(a), (b)) with $D_{x} = 2$ leading to the most efficient
structure.

In order to capture the structural features emerging in the best synchronizable networks,
we define the Pearson product correlation ($L_{\mathrm{corr}}$) between the degrees of the mirror
nodes in the pair of the layers as the evolution progresses as;
\begin{equation}
L_{\mathrm{corr}} = \frac{ (\sum_{i} (k_i - \langle k \rangle)(\sum_{i} (K_i - \langle K\rangle)}
{[ (\sum_{i} (k_i - \langle k \rangle)^2(\sum_{i} (K_i - \langle K \rangle)^2]^{1/2}}, 
\nonumber
\end{equation}
where $k_{i}$ and $K_{i}$ are the degrees of $i^{th}$ node in the first and the second layers, respectively. The terms $\langle k \rangle$ and $\langle K \rangle$
denote the average degrees of the first and the second layers, respectively. $\overline{L}_{\mathrm{corr}}$ is the average over the $L_{\mathrm{corr}}$ values
of the population used in the GA. For lower values of $D_{x}$($<$ 0.5), the optimized networks do not reveal significant
changes in
$\overline{L}_{\mathrm{corr}}$ and remain close to zero, with small random fluctuations. With a further increase in $D_{x}$ values, the average correlation $\overline{L}_{\mathrm{corr}}$ decreases for the optimized networks (Fig.~\ref{FigERAv6}(c)). This can be considered to occur in order to minimize the value of maximal load over the interlayer links.

The reasons behind the
emerging behavior of $\overline{L}_{\mathrm{corr}}$ might be explained as follows. When the higher degree nodes in one layer are connected with the higher degree nodes of the
second layer and the interlayer coupling strength is high as well 
($D_{x}$ $>$ 1), there is a very high load on the interlayer links, making it to become a dominant factor for affecting 
the synchronizability of the network \cite{Nishikawa}. Whereas, in a situation
when the higher degrees in one layer are connected with the lower degree nodes in another layer, the value of the maximal load on the interlayer links becomes less, leading to a 
homogeneity of the loads on the interlayer links.
In order to achieve this homogeneity, at higher values of $D_{x}$, the optimized networks 
exhibit negative values of $\overline{L}_{\mathrm{corr}}$.
These results demonstrate an impact of the multiplexity on the maximization of the 
synchronizability as well as an impact of the weighted
couplings on the evolved network architecture.

Further, in order to demonstrate the impact of structural properties of the initially selected 
networks population on the structural properties of the optimized networks,
we present the results for the scale-free networks taken as the initial population. The scale-free networks are generated using Barabasi-Albert algorithm \cite{rev_barabasi}.
As observed for the ER networks, the GA approach successfully minimizes the $R$ values for 
various $D_{x}$ values for the scale-free networks as well. The value of $D_{x}$, 
for which $R$ takes a minimum value, also remains same (Fig.~\ref{FigSFAv6}(a), (b)). 
The $L_{corr}$ value for the initial networks population
is positive (Fig.~\ref{FigSFAv6}(c)) since the mirror
nodes in both the layers are associated with the same time in the growth algorithm, 
but the optimized networks exhibit negative $L_{corr}$ values as $D_{x}$ increases (Fig.~\ref{FigSFAv6}(c)).
The efficiency of optimization for scale-free networks taken as the initial population, 
captured through $R_{norm}$, also exhibits a behavior similar to the ER random networks 
(Fig.~\ref{FigERAv6Norm}(b)) at the same interlayer coupling strength, 
though the values of $R_{norm}$
are lesser than those of ER random networks probably due to the higher values of $R$ for 
scale-free networks as compared to the ER random networks. 

\begin{figure}[t]
\centerline{\includegraphics[width=1.0\columnwidth]{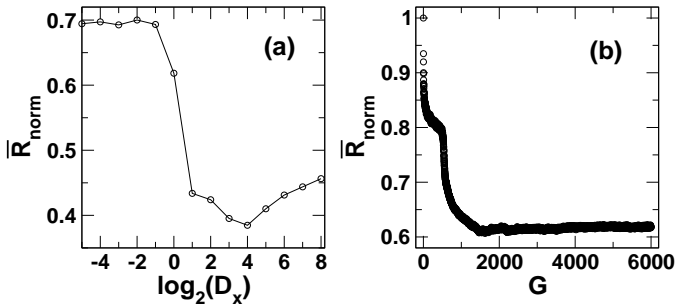}}
\caption{(a) Average values of $R_{norm}$ as a function of $D_{x}$. 
For each case GA minimize the $R_{norm}$ till 6000 iterations and the average of is taken over population of 1000 networks
and last 1000 iterations.
Panel (b) shows the convergence of average of $R_{norm}$ values over GA networks populations with generations(G) for $D_{x}$
= 1.
For both cases $E_{y}$ = 4 and each initial layer of ER networks have $\langle k\rangle$=6 and $N=100$.}
\label{Case}
\end{figure}
In order to analyze the impact of intralayer coupling strength on the global synchronizability, we introduce a parameter $E_y$. The new weighted adjacency matrix $M^i$ for the multiplex network can be defined as,
\begin{equation}
 M^i = \left[
    \begin{array}{cc}
      A^i~~D_{x}I\\
      D_{x}I~~E_{y}B^i
    \end{array}
\right]
\label{InraEq}
\end{equation} 
We consider one layer having intralayer coupling strength stronger
than that of the other layer ($E_{y} = 4$ in Eq.~\ref{InraEq}).  
With an initial increase in $D_{x}$, the values $R_{\mathrm{norm}}$ remains 
unaffected.
After a certain value of $D_{x}$, $R_{\mathrm{norm}}$ decreases drastically and thereafter for a certain range of $D_{x}$ the rate of decrement becomes less.
After attaining a minimum value, $R_{\mathrm{norm}}$ increases again
(Fig.~5(a)). 
This behavior shares a close similarity to the one observed for $E_{y}$ = 1 (Eq. 3), 
the major difference being the shift in the minima of $R_{\mathrm{norm}}$  
towards right, appearing at $D_{x}$ = 16. The probable reason behind this
shift might be rescaling of the $D_{x}$ values with respect to
$E_{y}$ towards the regime $D_{x}$ $<<$ 1. As a result, Eq. 3 holds good 
instead of Eq. 4 for the rescaled values of $D_{x}$ lying in the regime $D_{x}$ $<<$ 1.

Furthermore, there is a nearly similar behavior
of convergence of $R_{norm}$
with an increase in generation (G) of the GA. We find that initially with an
increase in the generation, the
$\overline{R}_{norm}$ values exhibit a faster decrement (Fig. 5(b)). After a certain value of G, the $\overline{R}_{norm}$ converges to a fixed value and the properties affecting the optimization of the networks population saturate. This leads to a steady value of $\overline{R}_{norm}$ which remains unaffected by a further increase in G.

\begin{figure}[t]
\centerline{\includegraphics[width=1\columnwidth]{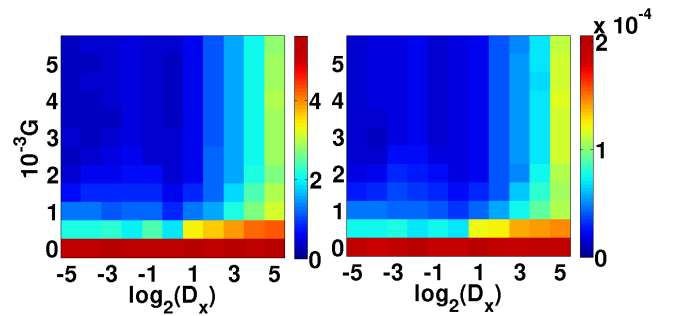}}
\caption{Average value of the variance of degree and betweenness centrality of
the nodes in the optimized networks in the left and right panels, respectively.
Initial layer of each ER network has $\langle k\rangle$=6 and $N=100$.}
\label{Av6Deg_BetAdj}
\end{figure}

Further, to analyze the  properties in the optimized networks, we calculate the variance in 
the degree and the betweenness centrality \cite{Newman_book}
for the different $D_{x}$ values. 
The initial ER random networks population has certain amount of 
degree heterogeneity due to
the Poisson degree distribution \cite{rev_barabasi}. For the 
lower interlayer coupling strength ($D_{x} < 2$), the heterogeneity is suppressed 
(Fig.~ \ref{Av6Deg_BetAdj}).
With a further decrease in the $D_{x}$ values, the suppression in the degree heterogeneity remains 
invariant, beyond which the suppression decreases in the optimized networks 
(Fig.~ \ref{Av6Deg_BetAdj}(a)).
Note that there is a higher heterogeneity in the connectivity for $D_{x}$ = 2 as compared to $D_{x}$ = 1.
Further, the betweenness centrality of a node can also be treated as a measure of the load 
distribution over the nodes in a network \cite{Luca}.
The ER random networks forming the initial population
again display a certain amount of the betweenness centrality heterogeneity. 
For the higher coupling strength, the heterogeneity in
betweenness centrality is not much suppressed in the optimized networks, however with 
a decrease in the interlayer coupling strength the suppression increases and 
converges to the homogeneous distribution at the lower coupling regime 
(Fig.~ \ref{Av6Deg_BetAdj}(b)). Though the optimized networks suppress heterogeneity in the 
degree as well as in the betweenness centrality, the interlayer coupling strength plays an 
instrumental role in deciding the amount of heterogeneity. The larger interlayer coupling 
strengths lead to more heterogeneity in the various structural properties of 
the optimized networks. 

These behaviors can be explained from the nature of Eq.~\ref{EQ1} and Eq.~\ref{EQ2}.
From Eq.~\ref{EQ1}, which holds good in the regime $D_{x}$ $>>$ 1, a structural change leads 
to a variation in the maxima of $\lambda_{max}$ over layers
by keeping its denominator constant for a given $D_{x}$ value.
Since the upper bound of $\lambda_{max}$
is an increasing function of the maximum degree of a network \cite{Das},
a minimized value of $\lambda_{max}$ is achieved when the maxima of the degree of networks in both the layers are minimized. However in regime $D_{x}$ $>>$ 1,
average value of $\lambda_{2}$ of the Laplacian for the layers appear in the
denominator of Eq.~\ref{EQ2}
which is lower bounded by the inverse of the diameter of the network
multiplied with the total number of
connections \cite{Chung_book}, providing
 another degree of freedom. As a result
sufficient amount of the homogeneity does not emerge in the optimized networks. 

{\bf Conclusion:} To conclude, we study the behavior of synchronizability in the multiplex 
networks with respect to interlayer coupling
strength. Various bounds for $D_{x}$ provide an understanding to how interlayer coupling 
strength turns out to be a deciding factor for achieving the most synchronizable networks as well as for 
the efficiency of the optimization. The results presented here have two fold applications, 
one being the influence 
of the weights in coupling strength on the synchronizability and the other pertains to the 
importance of the multiplex network framework in capturing 
the enhancement in the synchronizability and the identification of the structural features 
whose interplay can help in designing the best synchronizable 
networks. The behavior of the synchronizability with respect to the 
interlayer coupling strength is robust against changes in the initial networks architecture. 
Based on the eigenvalue ratio, we argue that a particular interlayer coupling 
strength value
leads to the highest efficiency of the optimization. Interestingly, the dynamical behavior of 
individual layers has been demonstrated to be largely affected while using them
in the multiplex network framework. While the individual layers may not be favoring 
synchronizability, the evolved multiplex networks can be best synchronizable demonstrating the importance of multilayer framework.  
Additionally, the betweenness centrality of the edges being treated as the 
load distribution \cite{Luca} explains the emergence of the negative correlation in the mirror 
nodes in the best synchronizable networks.

The framework and analysis presented here can be extended further in identifying a 
particular layer of the multiplex network which can be targeted to achieve the best or the
worst synchronizable multiplex networks.  
Further, the approach presented in this Letter can be adopted to optimize other dynamical
properties \cite{transport} in multiplex networks having restrictions on  
wiring topologies \cite{satyam}.

{\bf Acknowledgements:} SJ thanks CSIR, Govt. of
India project grant (25(0205)/12/EMR-II) and DST, Govt. of India project grant
(EMR/2014/000368) for financial support. SKD acknowledges University Grants Commission, Govt. of India,
for the fellowship.

\end{document}